\def\half{\frac{1}{2}}
\def\ph{\phantom{-}}  
\def\mb#1{\mbox{\boldmath{$#1$}}}
\def\sgn{{\rm sgn}}
\def\eq#1{Eq.\,(\ref{#1})}
\documentclass[amsmath,amssymb,superscriptaddress,showkeys, showpacs]{revtex4}
\usepackage{amsmath}
\usepackage{graphicx}
\usepackage{amstext}
\usepackage{mathrsfs} 
\begin{document}
\hspace*{5 in}CUQM-147
\vspace{0.5 in}
\markboth{R.~L.~Hall \& P. Zorin}{Nodal theorems for the Dirac equation in $d\ge 1$ dimensions}

\title{Nodal theorems for the Dirac equation in $d\ge 1$ dimensions}

\author{Richard L. Hall}
\email{richard.hall@concordia.ca}
\affiliation{Department of Mathematics and Statistics, Concordia University,
1455 de Maisonneuve Boulevard West, Montr\'eal,
Qu\'ebec, Canada H3G 1M8}

\author{Petr~Zorin}
\email{petrzorin@yahoo.com}
\affiliation{Department of Mathematics and Statistics, Concordia University,
1455 de Maisonneuve Boulevard West, Montr\'eal,
Qu\'ebec, Canada H3G 1M8}
\begin{abstract} 
A single particle obeys the Dirac equation in $d \ge 1$ spatial dimensions and is bound by an attractive central monotone potential that vanishes at infinity. In one dimension, the potential is even, and monotone for $x\ge 0.$  The asymptotic behavior of the wave functions near the origin and at infinity are discussed. Nodal theorems are proven for the cases $d=1$ and $d > 1$, which specify the relationship between the numbers of nodes $n_1$ and $n_2$ in the upper and lower components of the Dirac spinor. For $d=1$, $n_2 = n_1 + 1,$ whereas for $d >1,$ $n_2 = n_1 +1$ if $k_d > 0,$ and $n_2 = n_1$ if  $k_d < 0,$ where $k_d = \tau(j + \frac{d-2}{2}),$ and $\tau = \pm 1.$ This work generalizes the classic results of Rose and Newton in 1951 for the case $d=3.$ Specific examples are presented with graphs, including Dirac spinor orbits $(\psi_1(r), \psi_2(r)), r \ge 0.$
\end{abstract}



\maketitle

\section{Introduction}
Rose and Newton \cite{rose,rose_book} proved a nodal theorem for a single fermion moving in three dimensions in an 
attractive central potential which was not too singular at the origin.  The Dirac spinor for this problem 
is constructed with two radial functions, $\psi_1$ and $\psi_2$, with respective numbers of nodes $n_1$ and $n_2.$ The nodal theorem is to the effect that if $\kappa >0,$ then $n_2 = n_1+1,$ and if $\kappa < 0,$ then $n_2 = n_1,$ where $\kappa = \pm(j+\half).$ It is the purpose of the present paper to generalize this nodal theorem to $d \ge 1$ dimensions. We find it convenient to consider the cases $d=1$ and $d >1$ separately, since the parities of the radial functions are important only for $d=1$: thus we arrive at two theorems. In one dimension we find $n_2 = n_1+1,$ and in $d>1$ dimensions we prove a theorem which corresponds to the Rose-Newton result with $\kappa$ replaced by $k_d = \tau(j + \frac{d-2}{2}),$ where $\tau = \pm 1.$  

It it always interesting to know {\it a priori} the structure 
of the solution to a problem. For want of suitable theorems, the nodal structures of bound-state solutions of the Dirac equation have often been tacitly assumed in the
recent derivations  of relativistic comparison theorems, to the 
effect that $V_1 \le V_2 \Rightarrow E_1 \le E_2.$ Earlier, such theorems were not expected since the Dirac Hamiltonian is not bounded below and consequently it is difficult to characterize it's discrete spectrum variationally: thus comparison theorems are not immediately evident. The first results \cite{hallct1} were for ground-state problems in $d=3$ dimensions where the nodal structure was given by the Rose-Newton theorem. This Dirac comparison theorem was later extended to the ground state in $d\ge 1$ dimensions by G. Chen \cite{chen1, chen2}. More recent results were based on monotonicity arguments and yielded comparison theorems for the excited states \cite{hallct2, hallct3}; subsequently, applications to atomic physics \cite{softcore} were considered. These theorems and their application required either the precise nodal structure of the ground state, or the knowledge that, for an interesting class of problems, it made sense to compare the eigenvalues of states with the same nodal structure; in particular, the states needed to be suitably labelled. From a practical perspective, we have found that knowing the nodal structure is very helpful in computing bound-state eigenvalues numerically (by shooting methods) and for plotting the graphs of the radial functions $\{\psi_1(r), \psi_2(r)\}$ or the corresponding `Dirac spinor orbits' 
$\mb{r}(t) = (\psi_1(t),\psi_2(t)), t\ge 0$.  Articles by Leviatan {\it et al} \cite{gs2, gs3} show how Dirac nodal structure is important for nuclear physics.

\section{One--dimensional case $d=1$}
A Dirac equation in one spatial dimension and in natural units $\hbar=c=1$ for a potential $V$ which is symmetric with respect to reflection i.e. $V(-x)=V(x)$ reads \cite{calog}: 
\begin{equation*}
\left(\sigma_1\frac{\partial}{\partial x}-(E-V)\sigma_3+m\right)\psi=0,
\end{equation*}
where $m$ is the mass of the particle, $\sigma_1$ and $\sigma_3$ are Pauli matrices, and the energy $E$ satisfies $-m<E<m$, \cite{Spectrumd11, Spectrumd12}. Taking the two-component Dirac spinor as $\psi=\left(\begin{array}{cc}u_1 \\ u_2\end{array}\right)$ the above matrix equation can be decomposed into a system of first-order linear differential equations \cite{Dombey, Qiong}:
\begin{equation}\label{d1l}
u_1'=-(E+m-V)u_2,
\end{equation}
\begin{equation}\label{d1r}
u_2'=\ph(E-m-V)u_1,
\end{equation}
where prime $'$ denotes the derivative with respect to $x$. For bound states, $u_1$ and $u_2$ satisfy the normalization condition
\begin{equation*}
(u_1,u_1) + (u_2,u_2) = \int\limits_{-\infty}^{\infty}(u_1^2 + u_2^2)dx = 1.
\end{equation*}
Now we can state the theorem concerning the number of nodes of the Dirac spinor components for the one-dimensional case.

\medskip
\noindent{\bf  Nodal Theorem in $d=1$ dimension:} ~~{\it We assume $V$ is a negative even potential that is monotone for $x\ge 0$ and satisfies 
\begin{equation*}
\lim_{x\to 0}V(x)=-\nu \qquad \text{and} \qquad \lim_{x\to \pm\infty}V(x)=0,
\end{equation*} 
where $\nu$ is a positive constant. Then $u_1$ and $u_2$ have definite and opposite parities. 
If $n_1$ and $n_2$ are, respectively, the numbers of nodes of the upper $u_1$ and lower $u_2$ components of the Dirac spinor, then
\begin{equation*}
n_2=n_1+1. 
\end{equation*}} 

\medskip
\noindent{\bf  Proof:}~~
We suppose that $\{u_1(x), u_2(x)\}$ is a solution of Eqs.(\ref{d1l},~\ref{d1r}). 
Since $V$ is an even function of $x$, then by direct substitution we find that $\{u_1(-x), -u_2(-x)\}$ and $\{-u_1(-x), u_2(-x)\}$ are also solutions of Eqs.(\ref{d1l},~\ref{d1r}) with the same energy $E$. Therefore, by using linear combinations, we see that the solutions $\{u_1(x), u_2(x)\}$ of Eqs.(\ref{d1l},~\ref{d1r}) can be chosen to have definite and necessarily opposite parities, that is to say, $u_1(x)$ is even and $u_2(x)$ is odd, or {\it vice versa}. Also the energy spectrum of bound states is non-degenerate: details of this can be found in Refs. \cite{Coutinho, Qiong}.   

Now we analyse the asymptotic behavior of $u_1$ and $u_2$. At positive or negative infinity, after some rearragements, the system of Eqs.(\ref{d1l},~\ref{d1r}) becomes asymptotically
\begin{equation*}
u_1''=(m^2-E^2)u_1,
\end{equation*}
\begin{equation*}
u_2''=(m^2-E^2)u_2.
\end{equation*}
Solutions at infinity are given by 
\begin{equation*}
u_1=b^+_1e^{-x\beta}, \qquad  u_2=b^+_2e^{-x\beta},
\end{equation*}
and at negative infinity
\begin{equation*}
u_1=b^-_1e^{x\beta}, \qquad  u_2=b^-_2e^{x\beta},
\end{equation*}
where $b^\pm_1$ and $b^\pm_2$ are constants of integration and $\beta=\sqrt{m^2-E^2}$. Substitution of these solutions into Eqs.(\ref{d1l},~\ref{d1r}) as $x\longrightarrow\pm\infty$ gives us the values for $b^\pm_2$
\begin{equation*}
b^+_2=\frac{b^+_1\beta}{m+E} \qquad\text{and}\qquad b^-_2=-\frac{b^-_1\beta}{m+E}.
\end{equation*}
Considering the following two limits:
\begin{equation*}
\lim_{x\to+\infty}\frac{u_2}{u_1}=\frac{\beta}{m+E}>0
\qquad\text{and}\qquad
\lim_{x\to-\infty}\frac{u_2}{u_1}=-\frac{\beta}{m+E}<0,
\end{equation*}
we conclude that at positive infinity $u_1$ and $u_2$ have to vanish with the same signs and with different signs at negative infinity. We note that physics does not favour $\infty$ over $-\infty$: $u_1$ and $u_2$ can be interchanged by choosing a different representation for the matrix spinor equation leading to the coupled equations Eqs.(\ref{d1l},~\ref{d1r}).

We now introduce the following functions
\begin{equation*}
W_1(x):=E+m-V(x)
\qquad\text{and}\qquad
W_2(x):=E-m-V(x),
\end{equation*}
in terms of which Eqs.(\ref{d1l},~\ref{d1r}) become
\begin{equation}\label{d3l}
u'_1=-W_1u_2,
\end{equation} 
\begin{equation}\label{d3r}
u'_2=\ph W_2u_1.
\end{equation}
Since $V$ is even, and $u_1$ and $u_2$ have definite and opposite parities, for the remainder of the proof we restrict our attention to the positive half axis $x\ge 0$.
As $V$ is negative, and $-m<E<m$, it follows that $W_1>0$, $\forall x$. Also $V$ is monotone nondecreasing and vanishes for large $x$. Thus $W_2$ is monotone nonincreasing and $\lim\limits_{x\to\infty}W_2(x)=E-m<0$, but for small $x$, depending on $V$, $W_2$ may be positive. Hence, $W_2$ is either always negative or is positive near the origin and then changes sign once, say at $x=x_c$. We shall prove now that the wave functions $u_1$ and $u_2$ may have a nodes only on the interval $(0, x_c]$ where $W_2\ge 0$, and that these nodes are alternating. We observe parenthetically that, if $W_2$ is always negative, there are no excited states.
 
Let us start from the interval $(x_c, \infty)$ where $W_2<0$. Without loss of generality, we assume that at the point $x_1\in(x_c, \infty)$ 
the function $u_2$ is increasing and has a node, so $u_2(x_1)=0$ and $u_2'(x_1)>0$.
Then Eqs.(\ref{d3l},~\ref{d3r}) lead to
\begin{equation*}
u_1(x_1)<0, \quad u_1'(x_1)=0, \quad {\rm and } \quad u_1''(x_1)<0.
\end{equation*} 
Since $u_1$ and $u_2$ have to vanish at infinity, as $x$ increases  $u_2$ has to reach a local maximum at some point $x_2>x_1$ ($x_2$ is such that there are no nodes or extrema of $u_2$ between $x_1$ and $x_2$) and then decrease; thus $u_2(x_2)>0$, $u_2'(x_2)=0$, and $u_2''(x_2)<0$. From \eq{d3l} it follows that $u_1'(x_2)<0$ and from \eq{d3r} that $u_1'(x_2)>0$, which is not possible. A similar contradiction is reached if instead we consider a node of $u_1(x)$ at $x_1$. Thus nodes of $u_1$ and $u_2$ can occur only on $(0, x_c]$. 

Indeed, if we apply the same reasoning as above with $x_1$, $x_2\in (0, x_c]$, we find no contradiction; for $x>x_2$, the functions $u_1$ and $u_2$ can have more nodes or vanish as $x \longrightarrow \infty$. We also conclude that the nodes of $u_1$ and $u_2$ are alternating. The same conclusions are reached if $u_2$ is decreasing at $x_1$, or, as we observed above, if, instead of $u_2$, we consider $u_1$ first.

If we assume that $u_1$ is even and $u_2$ is odd, this leads to four cases at the origin:
\begin{equation*}
{\it case \ 1}: u_1(0)>0,\ u_1'(0)=0,\ u_2(0)=0,\ u_2'(0)>0, 
\end{equation*} 
\begin{equation*}
{\it case \ 2}: u_1(0)>0,\ u_1'(0)=0,\ u_2(0)=0,\ u_2'(0)<0, 
\end{equation*}
\begin{equation*}
{\it case \ 3}: u_1(0)<0,\ u_1'(0)=0,\ u_2(0)=0,\ u_2'(0)>0, 
\end{equation*}
\begin{equation*}
{\it case \ 4}: u_1(0)<0,\ u_1'(0)=0,\ u_2(0)=0,\ u_2'(0)<0. 
\end{equation*}
We can reduce these four cases to two because {\it case \ 1} and {\it case \ 2} correspond to {\it case \ 4} and {\it case \ 3}, respectively, multiplied by $-1$. We choose to keep {\it cases \ 1} and {\it 2}. Now, at $x=0$, $u_1$ is positive and has a local maximum. Eqs.(\ref{d3l},~\ref{d3r}) then imply: $u_2(0)=0$ and $u_2'(0)>0$, which rules out {\it case 2}. Thus, for $u_1$ even and $u_2$ odd,  we need only consider {\it case 1}.
  
Finally, {\it case \ 1} (and {\it case \ 4}) describe the behavior of $u_1$ and $u_2$ at the origin; the fact that the nodes of these components are alternating tells us how they behave further out, and we know that at infinity the components vanish with the same signs. Also we have assumed that  $u_1$ is even, so $u_2$ is odd, and reflection of their graphs gives us their behavior on $(-\infty,\ 0]$. In particular, $u_1$ and $u_2$ vanish at negative infinity with different signs. From the above it follows that $u_2$ has one node more than $u_1$. If $u_1$ is odd, then $u_2$ is even, and, by a similar analysis, we again find that $n_2 = n_1 +1.$ We observe that although $u_1$ can be node free when $u_2$ is odd, when $u_1$ is odd, $u_2$ must have at least two nodes. This completes the proof of the theorem.       
\hfill $\Box$

As an illustration for this result we plot wave functions for the laser-dressed Coulomb potential which is studied, for example, in Refs. \cite{laser1,laser2,Hall3}, and is given by
\begin{equation*}
V=-\frac{v}{(x^2+\lambda^2)^{1/2}}.
\end{equation*}
We consider $v=0.9$, $\lambda=0.5$, and $m=1$: in Figure~1, $u_1$ is even; and in Figure~2, $u_1$ is odd.
\begin{figure}
\centering{\includegraphics[height=13cm,width=5cm,angle=-90]{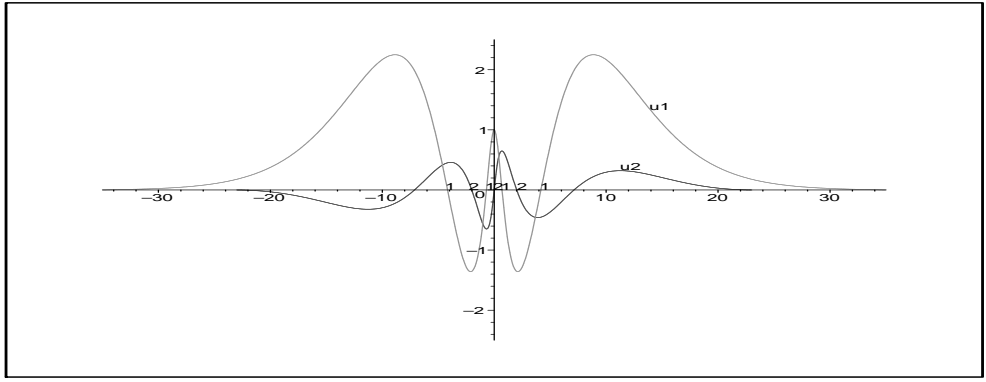}}
\caption{Dirac wave functions $u_1$ and $u_2$, such that $u_1$ is even with $u_1(0)>0$, $u_1'(0)=0$, and $u_2$ is odd with $u_2(0)=0$, $u_2'(0)>0$; $n_1=4$, $n_2=5$, $E=0.93011$.}
\end{figure}
\begin{figure}
\centering{\includegraphics[height=13cm,width=5cm,angle=-90]{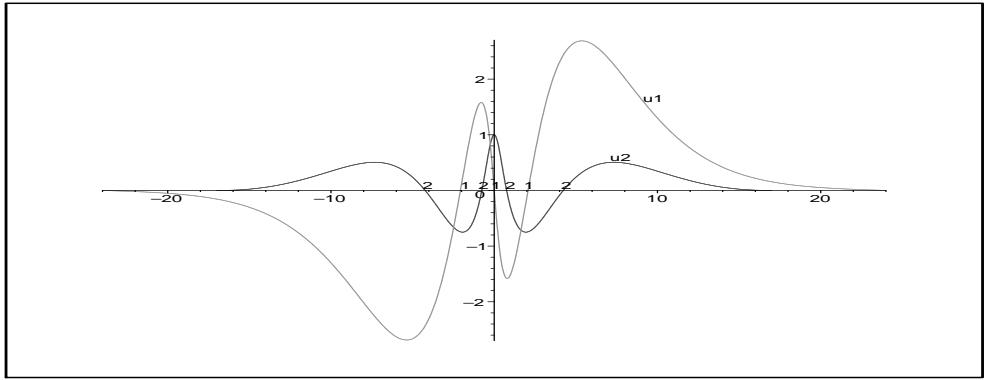}}
\caption{Dirac wave functions $u_1$ and $u_2$, such that $u_1$ is odd with $u_1(0)=0$, $u_1'(0)<0$, and $u_2$ is even with $u_2(0)>0$, $u_2'(0)=0$; $n_1=3$, $n_2=4$, $E=0.89177$.}
\end{figure}

\section{The higher dimensional cases $d>1$}
For a central potential in $d>1$ dimensions the Dirac equation can be written \cite{jiang} in natural units $\hbar=c=1$ as
\begin{equation*}
i{{\partial \Psi}\over{\partial t}} =H\Psi,\quad {\rm where}\quad  H=\sum_{s=1}^{d}{\alpha_{s}p_{s}} + m\beta+V,
\end{equation*}
where $m$ is the mass of the particle, $V$ is a spherically symmetric potential, and $\{\alpha_{s}\}$ and $\beta$  are the Dirac matrices which satisfy anti-commutation relations; the identity matrix is implied after the potential $V$. For stationary states, some algebraic calculations in a suitable basis, the details of which may be found in Refs. \cite{jiang, agboola, salazar, yasuk}, lead to a pair of first-order linear differential equations in two radial functions $\{\psi_1, \psi_2\}$, namely
\begin{equation}\label{dcel1}
\psi_1'=(E+m-V)\psi_2-\frac{k_d}{r}\psi_1,
\end{equation}
\begin{equation}\label{dcer1}
\psi_2'=\frac{k_d}{r}\psi_2-(E-m-V)\psi_1,
\end{equation}
where $r = \|\mb{r}\|$, prime $'$ denotes the derivative with respect to $r$, $k_{d}=\tau(j+{{d-2}\over{2}})$, $\tau = \pm 1$, $j=1/2$, $3/2$, $5/2$, $\ldots$. It is shown in Refs. \cite{Spectrum1, Spectrum2, Spectrum3} that the energy for bound states lies in the interval $(-m, m)$. We note that the variable $\tau$ is sometimes written $\omega$, as, for example in the book by Messiah \cite{messiah}, and the radial functions are often written $\psi_1 = G$ and $\psi_2 = F,$ as in the book by Greiner \cite{greiner}. For $d > 1,$ these functions vanish at $r = 0$, and, for bound states, they may be normalized by the relation 
\begin{equation*}
(\psi_1,\psi_1) + (\psi_2,\psi_2) = \int\limits_0^{\infty}(\psi_1^2(r) + \psi_2^2(r))dr = 1.
\end{equation*}
We use inner products {\it without} the radial measure $r^{(d-1)}$ because the factor $r^{\frac{(d-1)}{2}}$ is already built in to each radial function.  We shall assume that the potential $V$ is such that there is a discrete eigenvalue $E$ and that Eqs.(\ref{dcel1},~\ref{dcer1}) are the eigenequations for the corresponding radial eigenstates. Now we can state the Nodal Theorem for $d>1$.

\medskip
\noindent{\bf Nodal Theorem in $d>1$ dimensions:} ~~{\it We assume $V$ is a negative nondecreasing potential which vanishes at infinity. If $n_1$ and $n_2$ are the numbers of nodes of the upper $\psi_1$ and lower $\psi_2$ components of the Dirac wave function, then:
\begin{equation*}
n_2=n_1+1 \quad \text{if}\quad k_d>0
\end{equation*} 
and
\begin{equation*}
n_2=n_1 \quad \text{if}\quad k_d<0.
\end{equation*}}

\section{Asymptotic behavior of the wave functions}
\subsection{Near the origin}
Here we shall prove that near the origin the two components of the Dirac wave function start with the same signs if $k_d>0$ and with the different signs if $k_d<0$. We rewrite the system of Eqs.(\ref{dcel1},~\ref{dcer1}) in the following way
\begin{equation}\label{dcel3}
\psi_1'=W_1\psi_2-\frac{k_d}{r}\psi_1,
\end{equation}
\begin{equation}\label{dcer3}
\psi_2'=\frac{k_d}{r}\psi_2-W_2\psi_1,
\end{equation}
where 
\begin{equation*}
W_1(r):=E+m-V(r) \quad \text{and} \quad W_2(r):=E-m-V(r).
\end{equation*}
Since $-m<E<m$ and the potential $V$ is negative, the function $W_1>0$. Meanwhile, $W_2(r)$ is a monotone nonincreasing function which can change sign at most once, and if so from positive to negative; for reasons which will be clear later in section V we consider now only a region where $W_2\ge 0$. 

We first assume that $k_d>0$. Since $\psi_1(0)=0$, if $\psi_1\ge 0$ near the origin, clearly $\psi_1'\ge 0$. Then Eq.(\ref{dcel3}) implies that $\psi_2\ge 0$. Similarly, if $\psi_1\le 0$, near $r=0$, then $\psi_1'\le 0$ and it follows from Eq.(\ref{dcel3}) that $\psi_2\le 0$.

In the case $k_d<0$, we put $\psi_2\ge 0$ near the origin, then $\psi_2'$ has to be nonnegative as well and Eq.(\ref{dcer3}) leads to $\psi_1\le 0$. Finally, if $\psi_2\le 0$, it follows that $\psi_2'\le 0$ and Eq.(\ref{dcer3}) gives $\psi_1\ge 0$. Consequently, near the origin the wave functions $\psi_1$ and $\psi_2$ have the same signs if $k_d>0$ and opposite signs if $k_d<0$.

\subsection{At infinity}
In that section we shall prove that the two components of the wave function vanish at infinity with different signs. Since $\lim\limits_{r\to\infty}V(r)=0$, at infinity, Dirac coupled equations Eqs.(\ref{dcel1},~\ref{dcer1}) become
\begin{equation*}
\psi_1'=\ph(E+m)\psi_2,
\end{equation*}
\begin{equation*}
\psi_2'=-(E-m)\psi_1.
\end{equation*}
We know that the bound-state wave functions must vanish at infinity, so if $\psi_1\ge 0$ before it vanishes, then $\psi_1'\le 0$, and if $\psi_1\le 0$, then $\psi_1'\ge 0$. Now if we assume $\psi_1\ge 0$, the above equations lead to $\psi_2\le 0$ and {\it vice versa}. 
This means that at infinity, the component functions $\psi_1$ and $\psi_2$ vanish with different signs. This feature does not depend on the sign of $k_d$.
\begin{figure}
\centering{\includegraphics[height=11cm,width=5cm,angle=-90]{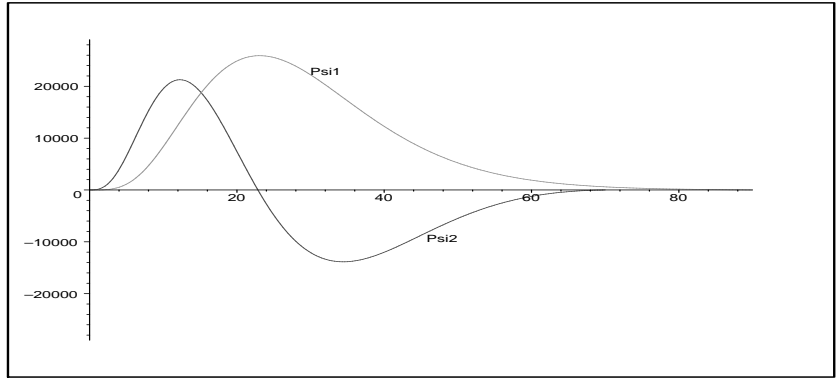}}
\caption{Wave functions have same signs near the origin and different at infinity: $n_1=0$, $n_2=1$, $\tau=1$, $E=0.98472$.}
\end{figure}
\begin{figure}
\centering{\includegraphics[height=11cm,width=5cm,angle=-90]{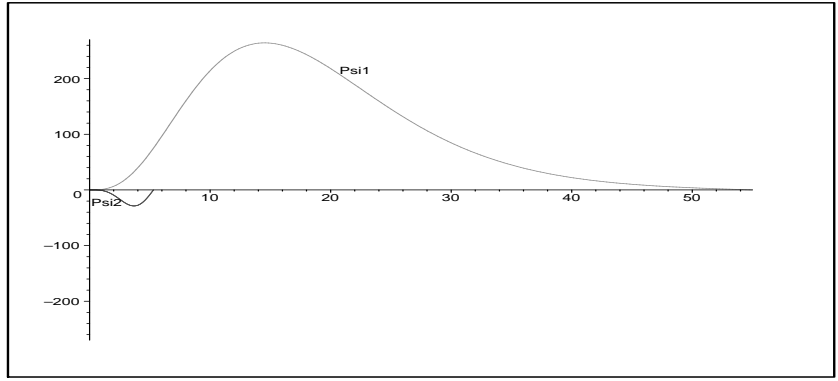}}
\caption{Wave functions have different signs near the origin and different at infinity: $n_1=n_2=0$, $\tau=-1$, $E=0.97487$.}
\end{figure}
\begin{figure}[ht]
\begin{center}
\begin{minipage}[ht]{0.4\linewidth}
\includegraphics[width=0.75\linewidth, angle=-90]{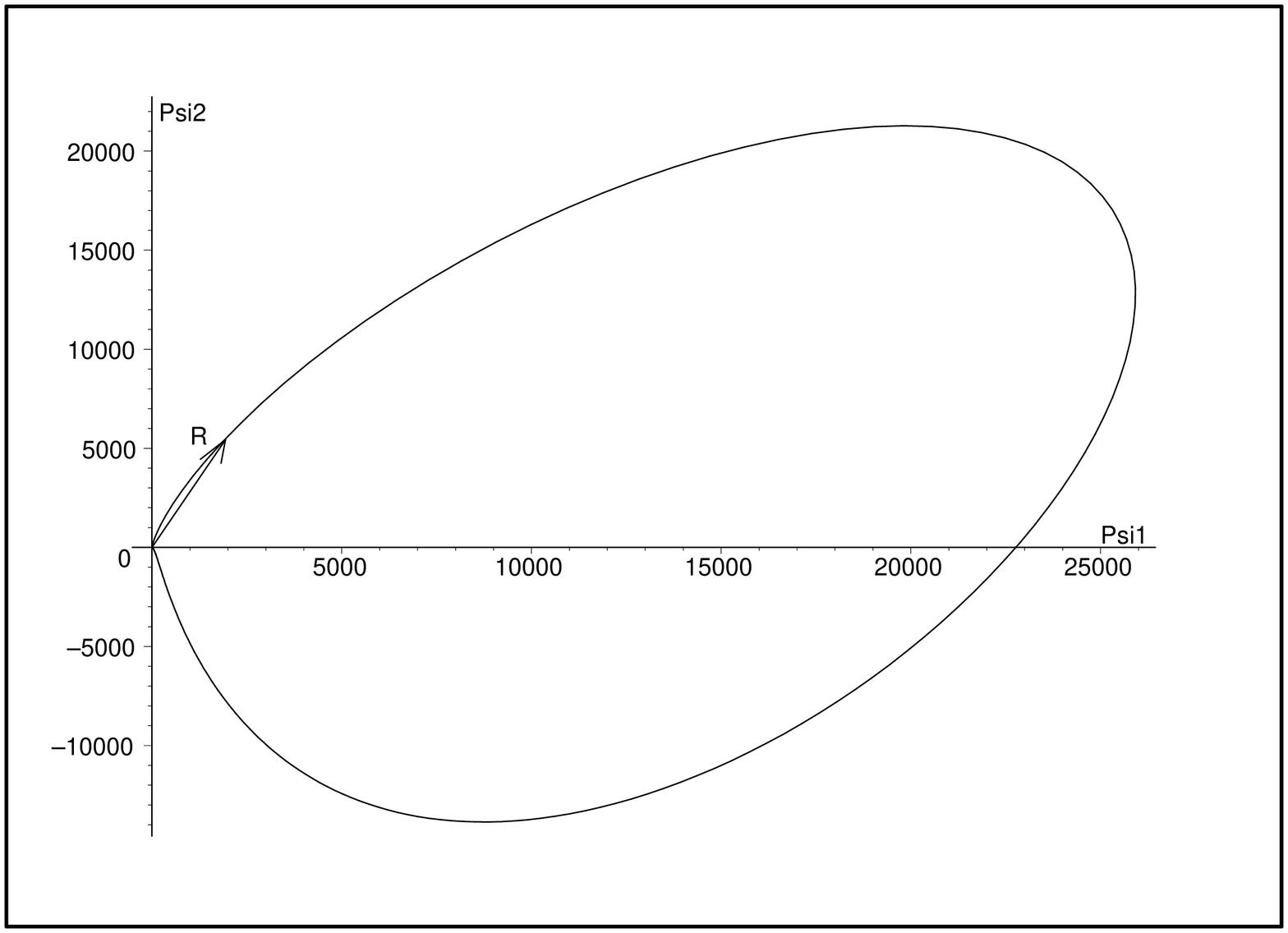}
\caption{Dirac spinor orbit corresponding to Fig. 3.} 
\end{minipage}
\hfill 
\begin{minipage}[ht]{0.4\linewidth}
\includegraphics[width=0.75\linewidth, angle=-90]{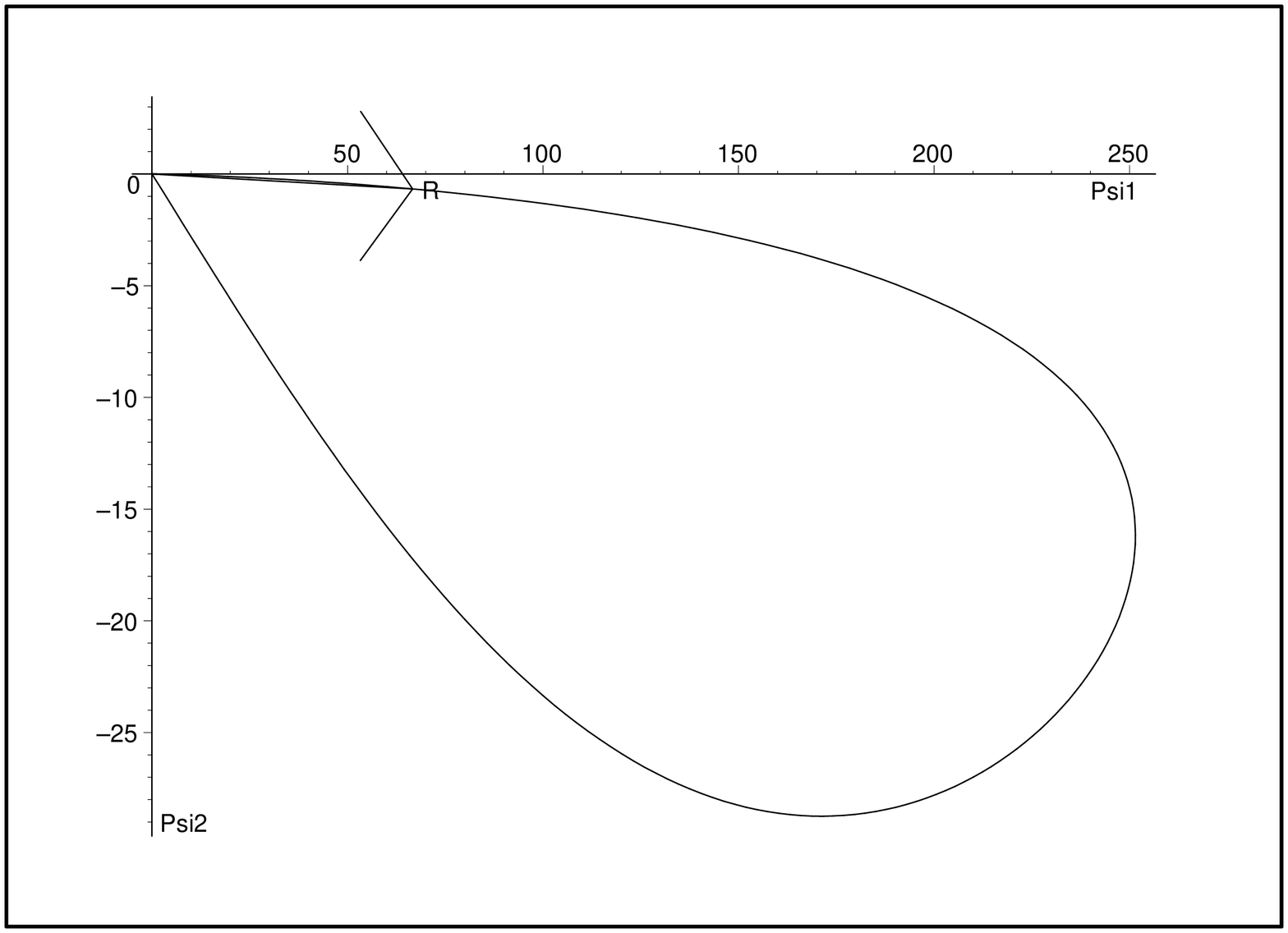}
\caption{Dirac spinor orbit corresponding to Fig. 4.}
\end{minipage}
\end{center}
\end{figure}

Now we study the graphs of the wave functions. 
We use Figures 3 to 10 as illustrations of the Nodal Theorem for $d>1$. We plot upper and lower components of the Dirac wave function, $\psi_1$ and $10\psi_2$ respectively, corresponding to the Hellmann potential \cite{hel1,hel2,hel3,Hall2} in the form $V(r)=-A/r+Be^{-Cr}/r$, with $A=0.7$, $B=0.5$, $C=0.25$, $m=1$, $j=3/2$, $d=5$, and $\tau = \pm 1$. If $k_d>0$ (Figure 3) 
both wavefunction components start at the origin with the same sign, but at infinity they must have different signs: thus one of the wave function components will have at least one node more than the other.
When $k_d<0$ (Figure 4) $\psi_1$ and $\psi_2$ start with different signs and then vanish with different signs, thus the numbers of nodes in $\psi_1$ and $\psi_2$ must be the same or differ by an even number. We can see from this easiest possible example that the number of nodes of the Dirac wave functions depends on the sign of $k_d$. These results are generalized by the Nodal Theorem for $d>1$.

\section{Proof of the Nodal Theorem in $d>1$ dimensions}
Now we let $u_1=r^{k_d}\psi_1$ and $u_2=r^{-k_d}\psi_2$. On the interval $(0, \infty)$, the nodes of $u_1$ and $u_2$ coincide with the nodes of $\psi_1$ and $\psi_2$, and Eqs.(\ref{dcel1},~\ref{dcer1}) in terms of $u_1$ and $u_2$ become
\begin{equation}\label{dcel2}
u_1'=\ph r^{2k_d}W_1u_2,
\end{equation}
\begin{equation}\label{dcer2}
u_2'=-r^{-2k_d}W_2u_1,
\end{equation} 
where 
\begin{equation*}
W_1(r):=E+m-V(r) \quad \text{and} \quad W_2(r):=E-m-V(r).
\end{equation*}
We immediately have that $W_1>0$ $\forall r\in[0, \infty)$. As in the one-dimensional case, we shall prove that the nodes of the radial wave functions $\psi_1$ and $\psi_2$ occur only when $W_2$ is nonnegative. Since $\lim\limits_{r\to\infty}V(r)=0$, and $E-m<0$, $W_2<0$ for $r$ sufficiently large.

Consider the interval $(r_c, \infty)$ on which $W_2<0$ and suppose that at a point $r_1\in (r_c, \infty)$ the function $u_2$ is increasing and has a node. Then at $r_1$, according to Eqs.(\ref{dcel2},~\ref{dcer2}), the function $u_1$ obeys the following conditions
\begin{equation*}
u_1'(r_1)=0, \quad u_1(r_1)>0, \quad {\rm and} \quad u_1''(r_1)>0.
\end{equation*}
Using the fact that $u_1$ and $u_2$ ($\psi_1$ and $\psi_2$ respectively) must both vanish at infinity we see that at some point $r_2>r_1$, such that there are no nodes or extrema of $u_2$ between $r_1$ and $r_2$, the function $u_2$ must reach a local maximum and then eventually decrease to zero. Therefore $u_2(r_2)>0$, $u_2'(r_2)=0$, $u_2''(r_2)<0$. \eq{dcel2} then implies $u_1'(r_2)>0$, but \eq{dcer2} leads to $u_1'(r_2)<0$, which is a contradiction. Thus the upper and lower components of the Dirac wave function intersect the $r$ axis only on $(0, x_c]$ where $W_2\ge 0$.

If we apply the same reasoning for $r_2\in(0, r_c]$, we find no contradiction, and moreover we conclude that the nodes of $\psi_1$ and $\psi_2$ are alternating. As in the one-dimensional case we arrive at the same results if $u_2$ is decreasing near $r_1$ or if we consider $u_1$ instead of $u_2$.

Now, following Rose and Newton \cite{rose} and using what we call `Dirac spinor orbits' \cite{Hallso}, we now consider the graphs generated parametrically by the points $(\psi_1(r), \psi_2(r))\in  \Re^2$, $r\ge 0$. We form a vector \mb{R} with initial point at the origin and terminal point $(\psi_1(r), \psi_2(r))$, $r\ge 0$, which makes an angle $\varphi$ with $\psi_1$ axis. Thus
\begin{equation*}
\rho=\tan\varphi=\frac{\psi_2}{\psi_1}.
\end{equation*}
From this follows
\begin{equation*}
\rho'=(1+\rho^2) \varphi'.
\end{equation*}
Since $(1+\rho^2)$ is positive then
\begin{equation}\label{sign}
\sgn(\rho')=\sgn(\varphi').
\end{equation}
Direct substitution of $\rho=\psi_2/\psi_1$ and equations Eqs.(\ref{dcel1},~\ref{dcer1}) shows us that $\rho$ satisfies the Ricatti equation
\begin{equation*}
\rho'=\frac{2k_d\rho}{r}-W_2-W_1\rho^2.
\end{equation*}
In the limit as $\rho$ approaches zero we have
\begin{equation*}
\rho'=-W_2\le 0.
\end{equation*}
Then if $\rho\longrightarrow\infty$, we have
\begin{equation*}
\rho'=-W_1\rho^2< 0,
\end{equation*}
since we consider only the region where nodes occur, i.e. $W_2\ge 0$.
Thus, from \eq{sign}, it follows that $\varphi'$ is a negative function of $r$ for $\rho\longrightarrow 0$ (or $\psi_2\longrightarrow 0$) and $\rho\longrightarrow\infty$ (or $\psi_1\longrightarrow 0$), which means that $\varphi$ is a decreasing function in the region $(0, r_c]$ where nodes of $\psi_1$ and $\psi_2$ occur. Therefore in the $\psi_1$--$\psi_2$ plane, the vector \mb{R} rotates clockwise.
\begin{figure}
\centering{\includegraphics[height=11cm,width=5cm,angle=-90]{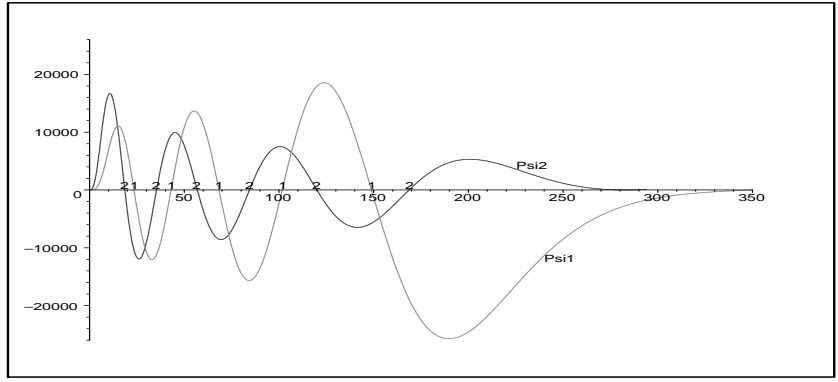}}
\caption{Wave functions $\psi_1$ and $\psi_2$ have the same signs near the origin and different at infinity:  $n_1=5$, $n_2=6$, $\tau=1$, $E=0.99697$.}
\end{figure}
\begin{figure}
\centering{\includegraphics[height=11cm,width=5cm,angle=-90]{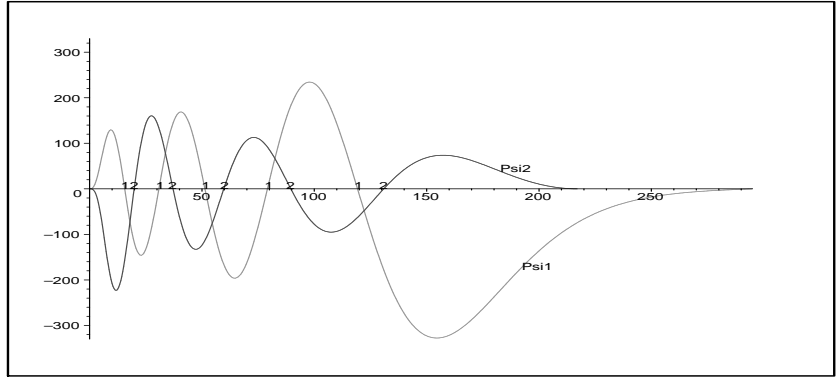}}
\caption{Wave functions $\psi_1$ and $\psi_2$ have different signs near the origin and different at infinity: $n_1=n_2=5$, $\tau=-1$, $E=0.99626$.}
\end{figure}
\begin{figure}[ht]
\begin{center}
\begin{minipage}[ht]{0.4\linewidth}
\includegraphics[width=0.75\linewidth, angle=-90]{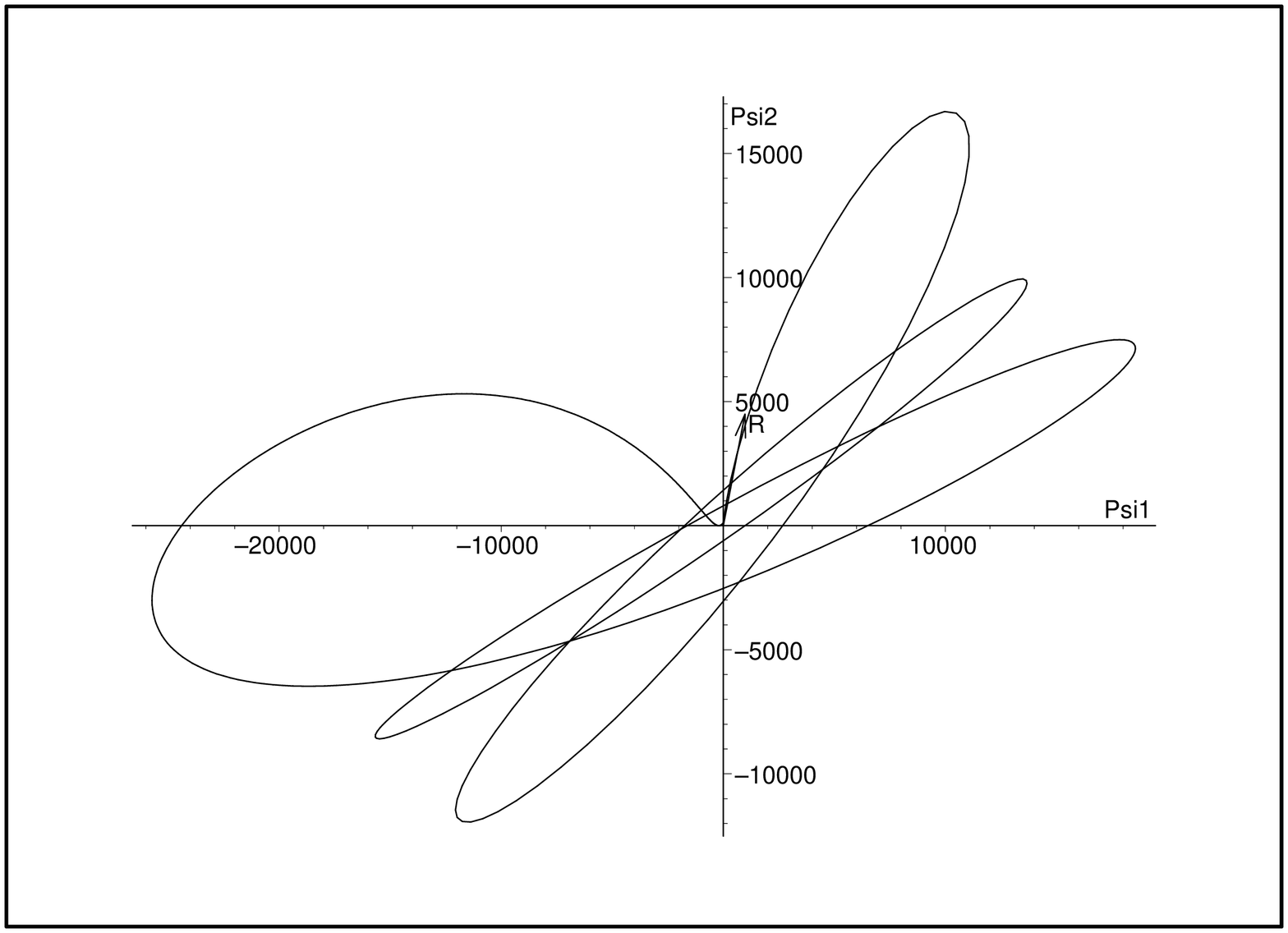}
\caption{Dirac spinor orbit corresponding to Fig. 7.} 
\end{minipage}
\hfill 
\begin{minipage}[ht]{0.4\linewidth}
\includegraphics[width=0.75\linewidth, angle=-90]{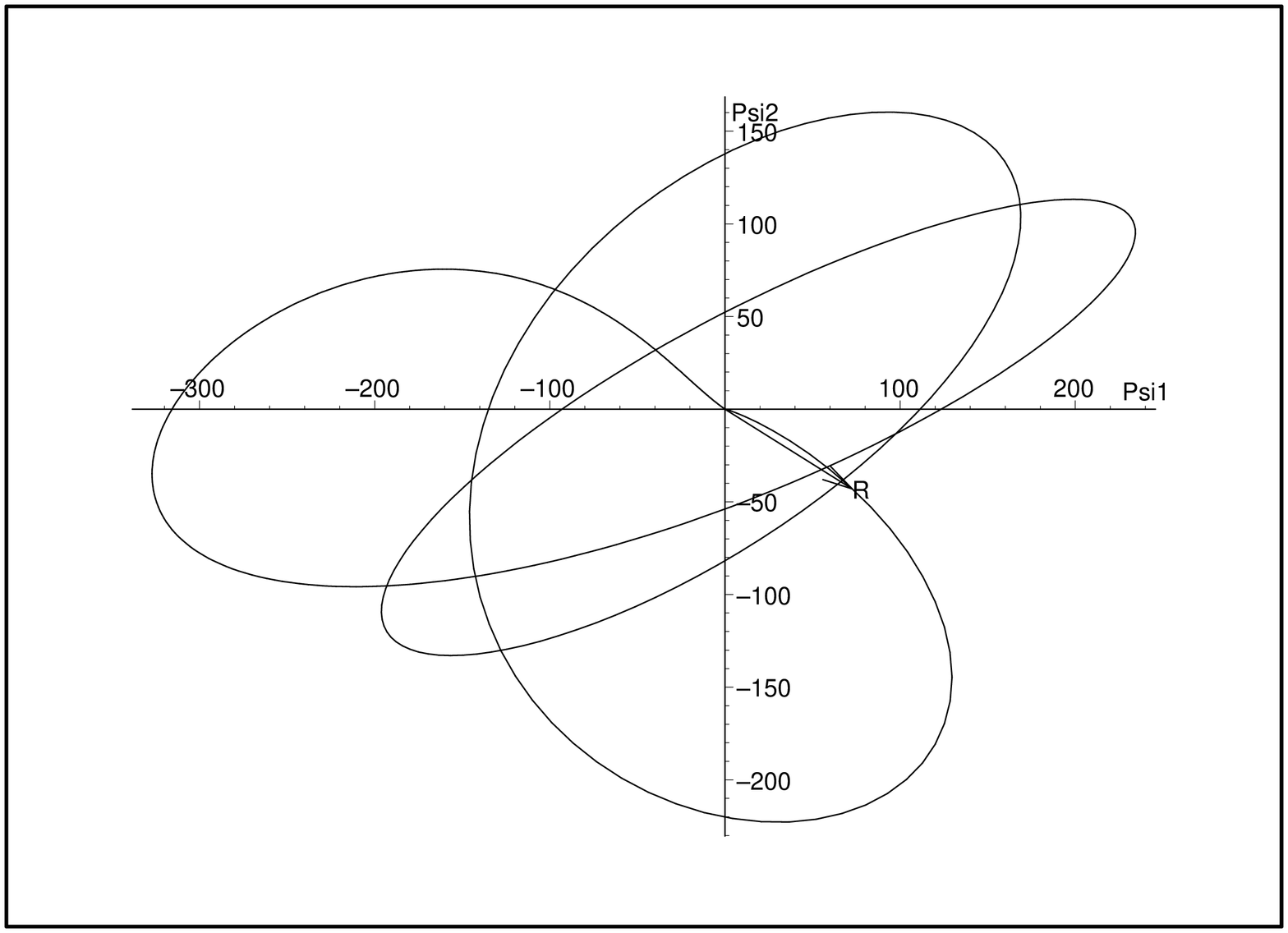}
\caption{Dirac spinor orbit corresponding to Fig.8.}
\end{minipage}
\end{center}
\end{figure}
Thus when $k_d$ and $\rho$ are positive (first and third quadrants), a node of $\psi_2$ occurs first; and when $k_d$ and $\rho$ are negative (second and fourth quadrants), a node of $\psi_1$ occurs first.

The case when $k_d>0$ is illustrated in Figure 7. Without loss of generality we consider the first quadrant, thus the wave functions start with the same positive sign and a node of $\psi_2$ occurs first; after this, $\psi_1$ has a node, and so on. At infinity $\psi_1$ and $\psi_2$ vanish with different signs. Therefore the number of nodes in the lower component of the Dirac wave function $\psi_2$ must exceed the number of nodes of $\psi_1$ by 1. Using the same reasoning for the case $k_d<0$, illustrated in Figure 8, we see that when $k_d<0$ the number of nodes in the upper and lower components of the Dirac wave function is the same. This completes the proof of the theorem.
\hfill $\Box$

We also plot the `Dirac spinor orbits', Figures 5, 6, 9, and 10, and they confirm the clockwise rotation of \mb{R}. We note that corresponding orbits can be plotted for $d=1$ case, but the vector \mb{R} would in this case rotate counterclockwise in the representation we have used.

\section{Conclusion}
At first sight a general nodal theorem for the Dirac equation in $d$ dimensions might appear to be out of reach. However, for central potentials, the bound states in all cases are built from just two radial functions.  For attractive potentials which are no more singular than Coulomb and which vanish at large distances, the logical possibilities are limited in a similar fashion to the three-dimensional case first analysed by Rose and Newton in 1951. We are therefore able to establish nodal theorems for all dimensions $d \ge 1$. Such theorems are useful for the study and computation of bound-state solutions, and they are essential for the establishment of relativistic comparison theorems.

\section{Acknowledgments}
One of us (RLH) gratefully acknowledges partial financial support
of this research under Grant No.\ GP3438 from the Natural Sciences
and Engineering Research Council of Canada.\medskip

\section*{References}

\end{document}